\documentclass[conference]{IEEEtran}
\IEEEoverridecommandlockouts
\usepackage{cite}
\usepackage{url}
\usepackage[cmex10]{amsmath}
\usepackage{mdwmath}
\usepackage{mdwtab}
\usepackage{amsfonts,amsmath,color,amssymb,amsxtra,graphicx}
\usepackage{booktabs}
\usepackage{multirow}
\usepackage{comment}
\usepackage[caption=false,font=footnotesize]{subfig}

\usepackage{epstopdf}

\usepackage{comment}

\begin{document}


%

\title{srsLTE: An Open-Source Platform for LTE Evolution and Experimentation}
\author{
  \authorblockN{Ismael Gomez-Miguelez, Andres Garcia-Saavedra, Paul D. Sutton,\\Pablo Serrano, Cristina Cano, Douglas J. Leith}
  
\thanks{I. Gomez-Miguelez and P. D. Sutton are with Software Radio Systems Limited, Ireland.}
\thanks{A. Garcia-Saavedra is with NEC Labs Europe, Germany.}
\thanks{P. Serrano is with Dept. of Telematics Engineering, Universidad Carlos III de Madrid, Spain.}
\thanks{C. Cano and D. J. Leith are with School of Computer Science and Statistics, Trinity College Dublin, Ireland.}

\thanks{This work was supported by Science Foundation Ireland grants 13/IF/I2781 and 11/PI/11771.}
}
\maketitle


\begin{abstract}

Testbeds are essential for experimental evaluation as well as for product development. In the context of LTE networks, existing testbed platforms are limited either in functionality and/or extensibility or are too complex to modify and customise. In this work we present srsLTE, an open-source platform for LTE experimentation designed for maximum modularity and code reuse and fully compliant with LTE Release 8. We show the potential of the srsLTE library by extending the baseline code to allow LTE transmissions in the unlicensed bands and coexistence with WiFi. We also expand previous results on this emerging research area by showing how different vendor-specific mechanisms in WiFi cards might affect coexistence.  

\end{abstract}

\section{Introduction}

Testbeds are today an essential platform for experimental research and prototype development. They enable researchers to test, validate and assess the performance of new technologies for wireless networks. In the context of LTE, a testbed typically includes one or several User Equipments (UEs), one base station (eNodeB) and an Evolved Packet Core (EPC), although the latter may be minimal. Each of these components is provided up front by commercial off-the-shelf (COTS) hardware solutions. Though usually expensive, the performance is excellent and the functionality is extensively validated. 

Some research problems, however, require adding new features to the standard or modifying some of its parts. For instance, in the context of 5G research, several groups are exploring how new waveforms such as GFDM can fit in the current LTE resource grid and physical layer procedures. Another example is IoT, where tight power and budget constraints may require simplified waveforms and protocols to be introduced in the standard and the performance assessed in a real scenario. Other research problems require instrumentation, often of the entire network stack (PHY to IP) to measure the impact these metrics might have on the user application. Such metrics include for example, the channel Doppler spread, the interference, the number of HARQ retransmissions, the number of turbo decoder iterations or the power headroom. Such modifications, customizations or instrumentation in COTS hardware are extremely difficult, if not impossible, or prohibitively expensive. 

Software-Defined Radio (SDR) is a popular concept for implementing radio equipment in software, using low-cost general purpose computers and radio frontends. In recent years, it's gaining popularity as a tool to build close-to-reality testbeds for experimental research. If the researcher has access to the SDR application code, as it is the case with open source, the testbed can be easily modified and instrumenting the stack becomes as simple as pulling out the required metrics from the code. Whilst in terms of performance or capabilities, open source SDR testbeds typically stand behind their commercial counterparts, this flexibility and openness can be much more valuable for many research problems. 

The  most popular open source LTE SDR software available for testbeds today are Eurecom's OpenAirInterface (OAI) \cite{oai} and openLTE \cite{openlte}. OAI currently provides a standard-compliant implementation of a subset of Release 10 LTE for UE, eNB, MME, HSS, SGw and PGw on standard Linux-based computing equipment (Intel x86 PC architectures). The software can be used in conjunction with standard RF laboratory equipment available in many labs (i.e. National Instruments/Ettus USRP and PXIe platforms). openLTE runs with the Ettus Research B2x0 USRP and provides eNB, MME and HSS  functionalities. openLTE code is well organized, documented and easy to customize or modify. However, it is incomplete and many features are still unstable or under development. Furthermore, it does not provide an UE, limiting the testbed capabilities in terms of instrumentation and measurement. OAI on the other hand is comparatively very complete and provides very good performance. However, the code structure is complex and difficult for a user external to the project to modify or customize.

In this work, we present an open source LTE library (srsLTE) and a complete software radio LTE UE (srsUE)~\cite{srslte}. We describe the architecture, evaluate the computational efficiency and discuss the suitability to research on future LTE enhancements.
As a case study we also present a modification to srsLTE that implements a duty cycle-based access mechanism for unlicensed LTE \cite{flore2014slides} and evaluate the results obtained in a coexistence scenario with WiFi. Given the attention that the use of the unlicensed spectrum by LTE is gathering, with concerns being raised by the Federal Communications Commission (FCC) \cite{FCC-note} as well as by the WiFi alliance \cite{WiFiAllianceStatement}, experimentation of emerging coexistence mechanisms in real testbeds becomes relevant in order to ensure that coexistence to WiFi is guaranteed in real deployments. Since the Medium Access Control (MAC) layer of LTE needs to be modified to implement changes in the way LTE accesses the channel, the srsLTE platform is a perfect candidate for experimentation. The results presented in this work extend those reported in \cite{jian2015coexistence}, that were obtained using a PHY-only implementation of LTE-A, by showing the impact of: \emph{i)} the LTE periodically leaving the channel empty to WiFi and \emph{ii)} vendor-specific nuances of different WiFi cards, which may affect WiFi performance in initially unforeseen ways.

The rest of this article is organised as follows. In Section \ref{sec:srsLTE} we describe the srsLTE library and analyse its computational efficiency, while the UE implementation is described in Section \ref{sec:srsUE}. Then, in Section \ref{sec:coexistence}, we present the LTE/WiFi coexistence scenario we use for evaluation and discuss the results obtained. Finally, we conclude with some final remarks.

\section{srsLTE: an Open-Source LTE Library for SDR}\label{sec:srsLTE}

In this section we describe the srsLTE library and we analyse its computational efficiency proving its suitability for experimentation with future enhancements of  the LTE standard.

\subsection{Description}

   \begin{figure*}[t!]
        \centering
             \subfloat[Module diagram for the srsLTE library.]{
	    \includegraphics[trim = 0.3in 0.2in 0.25in 0.18in, clip, width=0.5\linewidth]{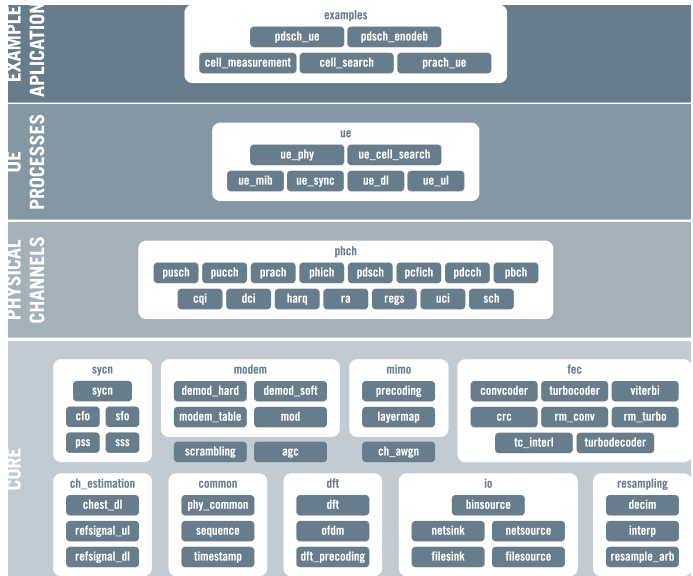}
	    \label{fig:class_diagram}
        }%
        \centering
        \subfloat[Threading architecture in srsUE. Boxes with coloured borders are threads.]{
	    \includegraphics[trim = 0.3in 0in 0.3in 0in, clip, width=0.5\linewidth]{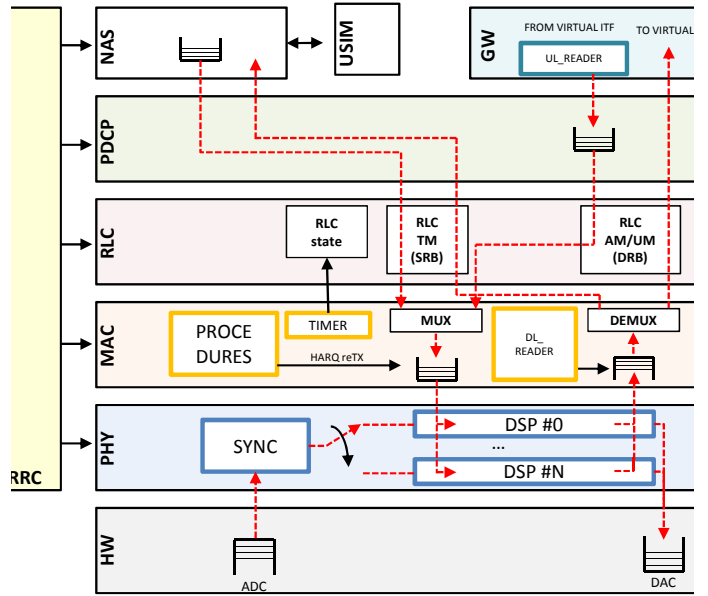}
	    \label{fig:layer_threading_arch}
        }      
  \caption{srsLTE library and srsUE threading architectures.}
  \label{fig:srsLTEsrsUE}
\end{figure*}

srsLTE is an open source library for the PHY layer of LTE Release 8. It is designed for maximum modularity and code reuse with minimal inter-module or external dependencies. The code is written in ANSI C and makes extensive use of Single Instruction Multiple Data (SIMD) operations, when available, for maximum performance. In terms of hardware, the library deals with buffers of samples in system memory thus being able to work with any RF front-end. It currently provides interfaces to the Universal Hardware Driver (UHD), giving support to the Ettus USRP family of devices. The aim of the library is providing the tools to build LTE-based applications such as a complete eNodeB or UE, an LTE sniffer or a network performance analyser.

The current features provided by the library are:

\begin{itemize}
\item LTE Release 8 compliant in FDD configuration;
\item Supported bandwidths: 1.4, 3, 5, 10 and 20 MHz;
\item Transmission mode 1 (single antenna) and 2 (transmit diversity);
\item Cell search and synchronization procedure for the UE;
\item All DL channels/signals are supported for UE and eNodeB side: PSS, SSS, PBCH, PCFICH, PHICH, PDCCH, PDSCH;
\item All UL channels/signals are supported for UE side: PRACH, PUSCH, PUCCH, SRS;
\item Highly optimized turbo decoder available in Intel SSE4.1/AVX (+100 Mbps) and standard C (+25 Mbps); \textit{and}
\item MATLAB and OCTAVE MEX library generation for many components.
\end{itemize}

The modular library approach allows researchers to easily customize, improve or completely replace components without affecting the rest of the code. Modules are organized hierarchically in the following categories, as illustrated in Figure \ref{fig:class_diagram}: 
\begin{itemize}

\item \textbf{Core}: The core modules are the main building blocks used within the PHY layer. In this category we find the turbo and convolutional coders and decoders, modulator and demodulator, synchronization, channel estimation and reference signal generation, OFDM and SC-FDMA processing and so forth. 
\item \textbf{Physical Channels}: There is one module for each uplink and downlink channel (e.g. PDSCH, PUSCH, PDCCH, PUCCH, etc). Each module uses the core building blocks to implement the signal processing required to convert bits into samples ready to send to the digital converter and vice versa. Some physical channels share some functionality, which is implemented in common auxiliary modules, e.g., PUSCH and PDSCH share many processing functions which are defined in the module SCH.
\item \textbf{UE Processes}: The UE processes implement the physical channel procedures for uplink and downlink making use of the physical channel modules. 
\item \textbf{Example Applications}: On top of the hierarchy we find a number of examples showing how to use the library through the UE processes modules. Among others, these examples include a PDSCH transmitter and receiver application and cell search examples. 
\end{itemize}

%

%

\subsection{Computational Efficiency}

Computational efficiency is the most challenging aspect of an SDR application. As in many other systems, the LTE receiver is much more complex than the transmitter. Though it can be ported to embedded processors, srsLTE initially targets general purpose processors (GPP), which is suitable for most research problems. Since memory is very cheap in GPPs, efficient code will make extensive use of look-up tables (LUT) and pre-generate as many signals as possible where memory access is more efficient than computation. For instance, all scrambling sequences, reference signals and some PUCCH signals can be generated for each subframe and input data combination. The CRC and encoder can use a LUT, and every interleaver can be pre-computed. 

Clock speeds for individual processing cores in GPPs are beginning to reach their practical limits. For this reason, further performance improvements in modern CPUs must exploit data and instruction parallelism. These techniques are used for the most expensive parts of the receiver, namely the turbo decoder, the channel estimator and equalizer, the demodulator and the Viterbi decoder. In order to provide maximum portability, the VOLK library \cite{volk} has been used as much as possible. VOLK is a library that contains kernels of hand-written SIMD code for different architectures, including a general one written in C. At runtime, VOLK will select the correct kernel for maximum performance. When not possible, compiler intrinsics are used to provide two alternative versions of some modules (a generic and a SIMD one) and the best one is chosen at compile time. 

We have measured the processing time of the UE PDSCH receiver for different bitrates (modulation and coding combinations) and analysed the per-module CPU usage in an Intel Core i7-3540M 3 GHz CPU. Table \ref{table:break_down} shows the obtained results. The PDSCH includes all symbol and bit processing, including OFDM demodulation, channel estimation, symbol detection, rate recovery, resetting the soft buffer, turbo decoding and CRC check. Since the interval at which data arrives in LTE is 1 ms, 1 processing core is able to process the entire PDSCH chain in real-time. In practice, more cores are needed because the CPU needs to perform other tasks, such as synchronization, uplink signal generation, physical layer procedures, etc. The per-module usage results indicate that the turbo decoder is by far the most demanding part and its relative weight in the system increases as the rate increases. The SIMD turbo decoder implementation in srsLTE is based on the max-log-MAP algorithm for SIMD introduced in \cite{turbo_decoder} with a minor modification to compute the horizontal maximum using the Intel's SSE4.1 \textit{phminposuw} instruction. 

\renewcommand{\arraystretch}{1.2}
\begin{table}
\caption{Total PDSCH receiver processing time and break-down of the CPU utilization for 20 Mhz bandwidth configuration. }
\begin{tabular}{c|c|c|c}

\toprule
\multirow{4}{*}{Module Name} & \multicolumn{3}{c}{Percentage of CPU} \\
 							 & 75 Mbps & 30 Mbps & 3.62 Mbps  \\							 
							 & 64QAM  & 16QAM  & QPSK  \\							 
\hline
Turbo decoder (1 iteration)	& 78.14 \%	& 64.21 \%	& 20.89 \%  \\
OFDM receive processing		& 6.08 \%	& 11.70 \%	& 33.33 \%  \\
Resource Element de-mapping	& 4.92 \%	& 9.31 \%	& 25.26 \%  \\
Rate recovery				& 4.49 \%	& 5.64 \%	& 8.34 \%  \\
CRC checksum				& 2.92 \%	& 2.23 \%	& 0.72 \%  \\
Soft demodulation			& 1.76 \%	& 2.11 \%	& 3.38 \%  \\
Equalization				& 0.16 \%	& 1.84 \%	& 4.98 \%  \\
Others						& 1.53 \%	& 2.96 \%	& 55.12 \%   \\
\bottomrule
\textbf{Total Execution Time} & 954 $\mu$s    & 488 $\mu$s    & 170 $\mu$s   \\ 
\bottomrule
\end{tabular}
\label{table:break_down}
\end{table}

\section{srsUE: A complete UE SDR Implementation}\label{sec:srsUE}

srsUE is a software radio LTE UE covering all layers of the network stack from PHY to IP. It is written in C++ and builds upon the srsLTE library which provides the PHY layer processing. For some security functions and RRC/NAS message parsing, it uses some functions from the openLTE project. Running on an Intel Core i7-4790, srsUE achieves more than 60 Mbps downlink with a 20 MHz bandwidth SISO configuration, when tested against an Amarisoft LTE 100 eNodeB. Apart from the features listed above for the srsLTE library, srsUE provides the following additional features: 
\begin{itemize}
\item MAC, RLC, PDCP, RRC, NAS and GW layers;
\item Soft USIM supporting Milenage and XOR authentication;
\item Detailed log system with per-layer log levels and hex dumps;
\item MAC layer Wireshark packet capture;
\item Command-line trace metrics;
\item Detailed input configuration file; \textit{and}
\item virtual network interface (i.e. \textit{tun/tap} device) created upon network attach.
\end{itemize}

A configuration file is provided to set parameters such as the downlink carrier frequency and log or packet capture options, making the software easy to use. The UE starts setting the USRP sampling rate to 1.96 MHz, in order to capture the synchronization PBCH signals. Once synchronization and MIB decoding is successful, it reconfigures the sampling rate to the appropriate sampling rate for the LTE signal bandwidth. Next, the UE attempts an attachment by sending a PRACH sequence and if the correct response is received, it continues the connection setup procedure. Following successful network attachment, a new virtual network interface is created in the system and the user can then establish IP sessions with the eNodeB network. 

This instrumentation and the Wireshark capture capabilities make srsUE an ideal tool for many applications, including cross-layer performance analysis, education and prototype development. srsUE opens the insights of the LTE stack to the user: the real-time traces, logs and packet captures can be used to correlate user experience in video streaming or web surfing with the signal quality or any other metric inside the stack, for example.

srsUE classes are organized by layers, one class per stack layer. Each class provides a separate clean interface to any other class that make use of it, which is used for message passing between layers of the stack.

A set of threads are created for performance and priority management reasons. Figure \ref{fig:layer_threading_arch} illustrates the different layers and the threads within them. Red dashed arrows indicate data paths whereas dark arrows indicate interaction between threads or classes. Each thread performs the following tasks:
\begin{itemize}
\item \textit{PHY DSP}: These threads perform all the processing associated with a subframe, including: OFDM demodulation, PDCCH search, PDSCH decoding, PUSCH/PUCCH encoding, uplink signal generation and transmission to the digital converter. After all processing is done, the thread returns to idle;
\item \textit{PHY SYNC}: Receives 1 subframe from the converter, performs time and frequency synchronization, copies the aligned frame into an idle PHY DSP thread and triggers its execution; 
\item \textit{MAC PROCEDURES}: Manages MAC procedures, including random access, scheduling request and uplink/downlink HARQ;
\item \textit{MAC TIMER}: Provides timer services to upper layers. All timers in LTE have a resolution of 1 ms, thus are implemented with simple counters synchronised to the subframe reception instead of using system clocks. 
\item \textit{MAC DL READER}: Reads downlink transport blocks from a buffer, processes the PDU and pushes the packet up the stack until the GW; \textit{and}
\item \textit{GW UL READER}: Reads from the virtual network interface and stores it into the PDCP buffer. 

\end{itemize}
 
The threading architecture in the PHY is motivated by the stringent latency requirements. In LTE, the required response time is 4 ms (4 subframes), which is imposed, in the UE, by (i) the transmission of an ACK/NACK HARQ indication after decoding a PDSCH transport block, (ii) the transmission of PUSCH after the reception of an uplink grant or (iii) the re-transmission of PUSCH after the reception of a NACK HARQ indication. The worst case scenario happens by the superposition of case (i) and (ii) or (i) and (iii): in the same subframe the UE decodes the PDSCH and encodes a PUSCH. Considering the time needed to buffer 1 subframe, the UE has 3 ms to decode the PDSCH, generate the PUSCH with the ACK/NACK information in it and send the samples to the converter buffer. 

Dividing the processing into uplink and downlink threads is ineffective, because the uplink processing needs to wait the downlink processing to finish. A more efficient approach is to parallelise all the processing associated with every subframe into a pipeline avoiding any delays (for thread synchronization or data transfer) within the tasks to be done in the 3 ms deadline. This architecture is efficient for dual- or quad-core CPUs. If more than 4 cores are available, increasing the number of PHY threads above 3 does not add any benefit, because the maximum tolerable latency is 3 ms (3 pipeline stages). The performance may then be improved by creating multiple threads associated to each subframe and split the multiple codeblocks in a transport block among them, a feature currently under development.

\section{Experimenting with an Unlicensed LTE \& WiFi coexistence scenario}\label{sec:coexistence}

As mentioned earlier, a potential use of the srsLTE platform is to analyse the coexistence of unlicensed LTE/WiFi.  The great advantage of using real hardware rather than simulations is that it allows us to evaluate the impact of the complex wireless propagation effects encountered indoors (where such small cells are most likely to be deployed) and also explore issues such as capture and carrier sense which are often difficult to model adequately.  

\subsection{LTE/WiFi Testbed}

Our interest is in a WiFi link and an LTE link operating in the same band. The LTE stations consist of an USRP~B210 board from Ettus connected via an USB~3.0 interface to a standard PC (Intel Core i7) running Linux Ubuntu Trusty, with the \texttt{uhd\_driver} and version 1.0.0 of srsLTE.  The distance from the LTE BS to the WiFi transmitter is 34~cm, and  35~cm to the WiFi receiver. The WiFi link is 94~cm long. We use a VERT2450 Dual Band (2.4 to 2.48 GHz and 4.9 to 5.9 GHz) omni-directional vertical antenna with 3dBi Gain and, unless otherwise noted, LTE is configured to use 100 physical resource block (PRBs), i.e., to use a 20~MHz channel bandwidth (the same as that used by WiFi). 

The WiFi nodes are Soekris net6501-70 devices, which are low-power single-board computers equipped with a 1.6~GHz Intel Atom E6xx series CPU, 2 Mini-PCI sockets, 2048 Mbyte DDR2-SDRAM and 8 Gbyte usb-based storage.  These run Linux Ubuntu (kernel 3.13). 
In all of our experiments, in order to detect vendor-specific performance issues that might be present in the 802.11 hardware (see e.g. \cite{bianchi}), we repeat all measurements using two different wireless NICs, namely, an Atheros~AR9390-based 802.11a/b/g/n card and a Broadcom BCM4321 card.  We configure the Atheros cards to operate in the 5~GHz band,  which measurements using a spectrum analyser confirmed were free from other transmissions, and configure the Broadcom cards to operate in the more populated 2.4~GHz band (since they do not support 5 GHz operation). 

\subsection{Impact of duty cycle}

\begin{figure}[t!]
\centering
\includegraphics[width=\linewidth]{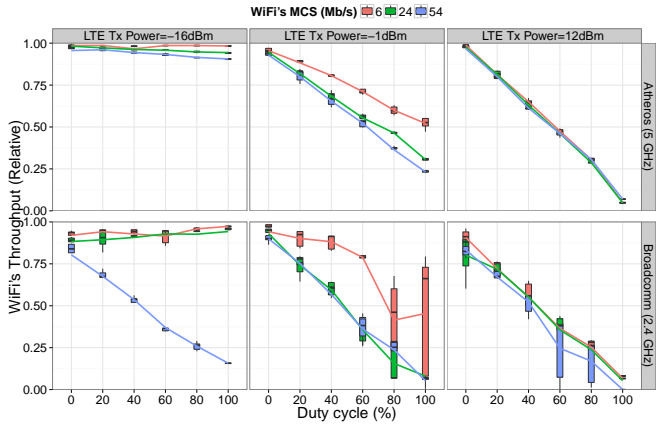}
\caption{Impact of LTE duty cycle on performance.}
\label{fig-duty_cycle}
\end{figure}

We start our experiments by investigating the impact of LTE transmissions on a WiFi link. To this end, we configured a  WiFi station to send 1500~Bytes UDP traffic.  The WiFi station is  constantly backlogged (always has a packet to send) and is configured to use a transmit power of 17~dBm.  We vary the activity on the LTE link by modifying the \texttt{srsLTE} code (namely, the example program \texttt{src/examples/pdsch\_enodeb.c}) that is executed in user space.  In more detail, we implement a periodic duty-cycle channel access scheme similar to the proposed CSAT coexistence mechanism \cite{sadek2015extending} where we fix an active interval during which the LTE base station transmits data to the UE followed by a randomized silent period.\footnote{Randomising the LTE silent period alleviates potential quantisation effects~\cite{cano2015unlicensed}.} The mean period (active plus silent) is equal to~150 ms.
By changing the mean duration of the silent period we vary the ``duty cycle'' of the LTE link. The LTE base station does not perform carrier sensing 
and sends traffic at the start of the frame boundary.  LTE control plane signals are transmitted during the on periods but the channel is completely freed during the silent periods.

We vary the LTE duty cycle between 0\% (no LTE transmissions) and 100\% (no LTE silent periods), and measure the resulting WiFi throughput for different settings of the LTE transmit power and of the Modulation and Coding Scheme (MCS) used by the WiFi link. The results are shown on Figure~\ref{fig-duty_cycle}, where for each configuration the average throughput is measured over a 10~s experiment, which is repeated 5 times, and we use box-and-whisker plots to represent the measured median, 25th and 75th percentiles, and the $1.5\times$~interquartile range. Note that we plot the normalised WiFi throughput, i.e., the ratio of the measured throughput to the maximum achievable throughput (with 0\% duty cycle), for ease of comparison across different choices of MCS.

In all of these experiments the WiFi throughput is inversely proportional to the LTE duty cycle, as might be expected as the LTE active and silent periods are relatively long compared to the WiFi transmissions (i.e. the LTE silent period leaves enough idle time to accommodate transmission of several WiFi packets).  When the 5~GHz band is used the results show little dispersion and the use of a lower MCS increases the robustness of the WiFi transmissions.  However, this effect is minor when the LTE transmission power is very low (so the LTE interference with WiFi is then negligible) and also when the LTE transmission power is high (when all choices of WiFi MCS are equally affected by the interference from LTE).  When the 2.4~GHz band is used the results are qualitatively similar, although there is more dispersion, 
 with the 54~Mbps MCS in particular being extremely sensitive to LTE interference.

To sum up, these measurements confirm that scheduling the LTE transmissions using a randomised duty cycle, with relative long on and off periods, has the expected qualitative impact for both of the WiFi wireless cards considered.   We next analyse in more detail the impact on performance of the choice of MCS, transmission power and band used.

\subsection{Impact of transmission power}

\begin{figure}[t!]
\centering
\includegraphics[width=\linewidth]{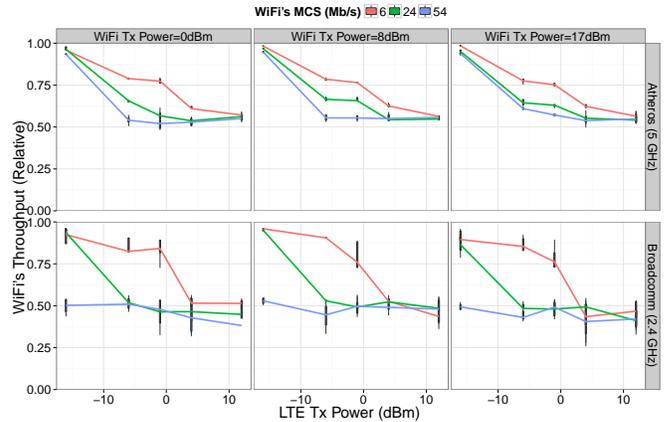}
\caption{WiFi throughput vs. LTE transmission power}
\label{fig-gain}
\end{figure}

The foregoing results confirm that adjusting the LTE duty cycle allows the impact of LTE interference on WiFi to be controlled. We now fix the LTE duty cycle to be 50\% and vary the transmission power of both the WiFi and LTE transmitters, with the aim of understanding the sensitivity and robustness of WiFi performance to these parameters. Measurements are shown in Figure~\ref{fig-gain}, for the two WiFi wireless cards studied and for different choices of WiFi MCS. 

It can be seen that the WiFi transmission power has little, if any, impact on performance.  This confirms that, in our small testbed, the signal quality of the WiFi link is good for all the configurations of this parameter. The measurements also show that, as in the previous experiments, the use of a lower MCS results in an increased WiFi normalised throughput, which suggests that a large number of WiFi transmissions collide with LTE transmissions (i.e., the reduction in throughput performance is not only caused by channel deferral), and therefore the use of more robust transmission schemes favors the appearance of the \emph{capture effect}. In fact, for the case of the Atheros card, there is a small but consistent improvement in performance as the transmit power is increased from 8~dBm to 17~dBm when the MCS is 54~Mbps and the LTE transmission power is -6~dBm, which lends further support to this observation. 

Given these results, we conclude that careful tuning of the transmission power could potentially be used to adjust the share of resources between WiFi and LTE.  However, given the steepness of the variation in throughput vs. LTE transmission power and the only minor impact of the WiFi configuration, this might not be a practical approach.

\subsection{Impact of bandwidth used}

\begin{figure}[t!]
\centering
\includegraphics[width=\linewidth]{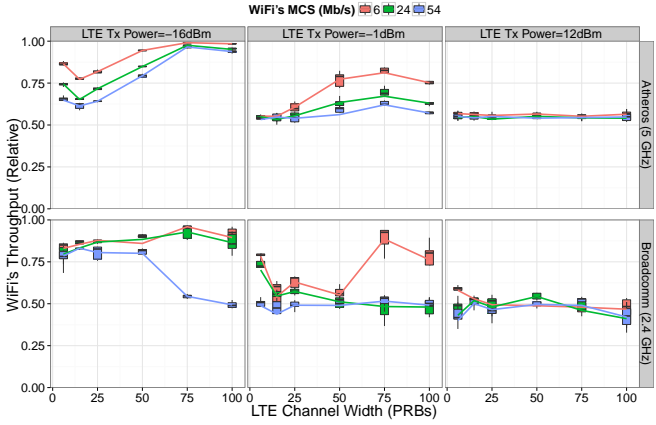}
\caption{WiFi throughput vs. LTE bandwidth}
\label{fig-prb}
\end{figure}

We next investigate the impact of varying the number of subcarriers used by LTE (i.e., its bandwidth). To this end, we perform a sweep on the number of PRBs used by LTE, and meausre the resulting WiFi throughput for different choices of the LTE transmission power and WiFi MCS.  The measurements obtained are shown in Figure~\ref{fig-prb}. 

In contrast to the previous experiments, here there are significant differences in the WiFi throughput depending on the WiFi card used.  While when the LTE transmit power is 12~dBm the number of PRBs does not affect WiFi throughput, when the LTE transmit power is -16~dBm the WiFi throughput varies considerably as the number of PRBs changes.  It can also be seen that the Atheros and Broadcom WiFi NICs exhibit the \emph{opposite} behaviour: for the latter, when the MCS is 54~Mbps the throughput reduces as interference increases (an arguably intuitive result, as LTE occupies more spectrum), while for the former throughput increases with the number of PRBs used. Furthermore, this behaviour is qualitatively consistent for the case of -1~dBm with the Atheros cards, while the Broadcom cards and an MCS of 6~Mbps now show a different pattern as compared to when the LTE transmit power is -16~dBm.  Our hypothesis for this behaviour is that these WiFi cards make use of different (proprietary) clear channel assessment mechanisms for carrier sensing.  

These results indicate that while changing the number of PRBs used by LTE might be useful to manage coexistence between LTE and WiFi, this may result in vendor-dependent performance.  One potential consequence is that this may introduce fairness issues in a multi-vendor environment.  Therefore, we argue that more experimental research, such as the enabled by our srsLTE-based platform, is required prior to the rollout of LTE in unlicensed bands.

\subsection{Impact of the central frequency}

\begin{figure}[t!]
\centering
\includegraphics[width=\linewidth]{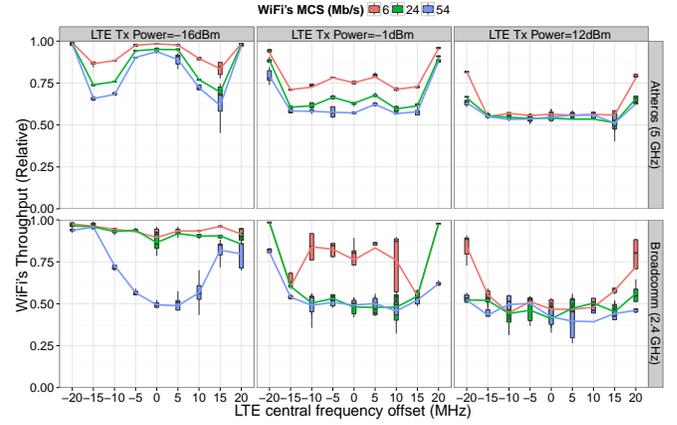}
\caption{WiFi throughput vs. LTE central frequency}
\label{fig-central_freq}
\end{figure}

Finally, we explore performance when the amount of spectrum overlap between LTE and WiFI is varied.  We again  configure LTE to use 100~PRBs (corresponding to a channel bandwidth of 20MHz) and now vary the central frequency used for LTE transmissions from -20~MHz to +20~MHz in steps of 5~MHz.  Similarly to the previous experiments, we perform the tests for a range of LTE transmission powers and WiFi MCS.  The resulting measurements are shown in Figure~\ref{fig-central_freq}. 

On the one hand, when the LTE transmission power is highest (12~dBm) then the WiFi throughput performance is similar for both wireless cards: when there is large overlap in the spectrum used by LTE and WiFi the interference is large and only for the edge cases where the centre frequency is $\pm$~20~MHz is it reduced (it is also worth noting that the interference pattern is practically symmetrical).  On the other hand, when the LTE transmission power is smallest (-16~dBm), the behaviour changes depending on the WiFi card used.  For the case of the Broadcom card the results are as might be expected: the smaller the frequency offset the lower the WiFi throughput for an MCS of 54~Mbps, while for more robust MCS the performance does not change significantly with the frequency offset.  For the case of the Atheros cards instead of a U-shaped figure we see that, for all choices of MCS, throughput is maximised either when the LTE transmissions are placed 20~MHz away from the center frequency or when LTE uses the exact same frequency as WiFi transmissions.   For intermediate LTE transmission powers (-1~dBm), although in all cases WiFi throughput improves when using a more robust MCS, the results vary qualitatively depending on the hardware considered: for the case of Atheros, the offset has little impact on thoughput apart from when it is 20~MHz; in contrast, for the Broadcom cards and an MCS of 6~Mbps there is a remarkable  drop in throughput for an offset of 15~MHz.  These measurements confirm the presence of a dependency on the WiFi hardware used when considering LTE/WiFi coexistence.

These results further support the two conclusions in the previous experiments: firstly, an adequate tuning of the spectrum used by LTE may improve coexistence with WiFi; secondly, more real-life experimentation is required to better understand the reasons for the observed behaviour and to detect vendor-specific issues that might preclude fair coexistence.

\section{Conclusions}

In this work we have presented an open-source, modular and fully compliant with LTE Release 8 platform that allows for LTE extension and experimentation. We have described the architecture of the srsLTE library as well as the srsUE and evaluated its computational efficiency proving its suitability to current LTE testing. As a case study we have shown the potential of the library for extension of LTE to work in the unlicensed bands and coexist with WiFi networks. Our results in this regard motivate further experimental validation of emerging coexistence mechanisms as particularities of off-the-shelf wireless cards might affect fair coexistence in ways difficult to predict via analysis or simulations. 

\bibliographystyle{IEEEtran}
\bibliography{bibliography}

\end{document}